\documentclass[12pt]{article}

\usepackage{graphicx}

\usepackage{geometry} 
\usepackage[cp1251]{inputenc}
\usepackage[english,russian,ukrainian]{babel}

\begin{document}

\textbf{Asymptotic of the electric structure function and the deuteron wave
function}

\begin{center}
\textbf{\textit{V. I. Zhaba}}
\end{center}

\begin{center}
\textit{Uzhgorod National University, Department of Theoretical
Physics,}
\end{center}

\begin{center}
\textit{54, Voloshyna St., Uzhgorod, UA-88000, Ukraine}
\end{center}

\begin{center}
\textit{E-mail: viktorzh@meta.ua}
\end{center}

\textbf{Abstract}

The main features of obtaining the asymptotic behaviour of the
electric structure function $A(p)$ at large values of the
transmitted momentum are analysed. The asymptotic behaviour of the
structure function $A(p)$ was determined to take into account the
asymptotic behaviour of the deuteron form factors and the original
dipole approximation for the nucleon form factors. Asymptotic
values of $A(p)$ were obtained for the nucleon-nucleon potential
Reid93 and compared with the calculations for different nucleon
form factors models and their approximations. In the broad
momentum range up to 12.5 fm$^{-1}$, the basic forms of the
asymptotic behaviour of the electric structure function are
demonstrated and compared with the experimental data of the modern
collaborations. As the analysis shows in most cases considered,
the asymptotic for $A(p)$ is represented in the form of the power
function $p^{-n}$.

\textbf{Абстракт}
Проаналізовано основні особливості одержання асимптотичної поведінки 
функції електричної структури $A(p)$ при великих значеннях переданого імпульсу. 
Асимптотична поведінка структурної функції $A(p)$ 
була визначена при врахуванні асимптотичної поведінки 
дейтронних формфакторів та оригінального дипольного наближення 
для нуклонних формфакторів. 
Для нуклон-нуклонного потенціалу Reid93 одержано 
асиптотичні значення $A(p)$ і порівняно їх із розрахунками 
для різних моделей нуклонних формфакторів та їх наближень. 
В широкому діапазоні імпульсів до 12.5 fm$^{-1}$ продемонстровано 
основні форми асимптотичної поведінки функції електричної структури 
та порівняно їх з експериментальними даними провідних колаборацій. 
Як показує аналіз, у більшості розглянутих випадків асимптотика 
для $A(p)$ представлена у формі степеневої функції $p^{-n}$.

\textbf{Keywords: }structure function, asymptotic, deuteron, form
factor, approximation, wave function.

PACS: 13.40.Gp, 13.88.+e, 21.45.Bc, 03.65.Nk

\textbf{1. Introduction}

As a two-nucleon bound state, the deuteron is the simplest system in nuclear
physics. This simple two-particle system allows us to better understand the
nucleon-nucleon interaction in the processes of deuteron scattering.
Measurements of the electromagnetic properties of a deuteron are invaluable
for studying the internal structure of a deuteron.

The electromagnetic structure of the deuteron is observed in the
elastic scattering of electrons on deuterons and can be described
by the three form factors $G_{C}$, $G_{Q}$ and $G_{M}$, which are
the electric monopole, the electric quadrupole and the magnetic
dipole distributions of the deuteron respectively in transferred
momentum squared $p^{2}$ representation \cite{Hasell2011}. From
measurements of unpolarized elastic electron-deuteron
cross-section at different scattering angles, two combinations of
the deuteron form factors $A(p)$ and $B(p)$ can be obtained (by
Rosenbluth separation).

The deuteron elastic structure function $A(p)$ can be extracted
from cross-section measurements of elastic scattering of electrons
on deuterons in coincidence (for example, in the range 0.7$ \le
p^{2} \le $6.0 (GeV/c)$^{2}$ for Hall A of Jefferson Laboratory
\cite{Alexa1999}). At low values of momentum $p^{2}$, the
cross-section is dominated by $A(p)$, and in $A$ is dominated by
form factor $G_{C}$ \cite{Hasell2011}. One of the latest
experimental measurements for the electric structure function
$A(p)$ was presented in an intermediate momentum transfer region
in \cite{Abbott1999} and for large momentum in \cite{Alexa1999,
PR08019}. The experimental status of $A(p)$ is described in detail
in the review \cite{Gilman2002}.

We note another important feature of the knowledge of the electric
structure function $A(p)$. It is a component in tensor $t_{2j}$
and vector $t_{1i}$ polarizations \cite{Gilman2002}, in the spin
correlation coefficients and tensor asymmetries \cite{Gakh2014},
and other polarization observables in processes with the
participation of deuteron. The value $A(p)$ appears in the factor
$S(p,\theta _e ) = A(p) + B(p)tg^2(\theta _e / 2)$, where
\textit{$\theta $}$_{e}$ is the electron scattering angle in the
laboratory system.

A theoretical studies of the value $A(p)$ in ed-scattering using
modern theoretical approaches and approximations remains relevant
\cite{Phillips2005, Kohl2008, Piarulli2013, Khokhlov2015,
Gutsche2016}. In this paper, we use the analytic forms of the
deuteron wave function (DWF) in coordinate representation for
theoretical calculations of electric structure function $A(p)$ and
its asymptotic at large momentums in the elastic electron-deuteron
scattering. The nucleon-nucleon realistic phenomenological
potential of Nijmegen group Reid93 \cite{Stoks1994, Swart1995,
MPLA3125} and the different models of nucleon form factors were
used for numerical calculations.

\textbf{2. The electric structure function}

Different models of NN potential for quantitative understanding of
the structure of the deuteron, S- and D-states and polarization
characteristics are considered. The deuteron charge distribution
is not well known from the experiment because it is determined
from the data of polarization experiments (polarizations and
differential cross-sections) \cite{Abbott20001}. Differential
cross-section of elastic scattering of unpolarized electrons by
unpolarized deuterons without measuring the polarization of
reflected electrons and deuterons is given by the formula within
the assumptions of the first Born approximation and the conditions
of relativistic invariance \cite{Elias1969, Galster1971,
Donnelly1986, Garcon2001, Gilman2002}

\begin{equation}
\label{eq1}
\frac{d\sigma }{d\Omega _e } = \left( {\frac{d\sigma }{d\Omega _e }}
\right)_{Mott} \left[ {A(p^2) + B(p^2)tg^2\left( {\frac{\theta _e }{2}}
\right)} \right].
\end{equation}

Formula (\ref{eq1}) is obtained by Rosenbluth
\cite{Rosenbluth1950}. Here $\left( {\frac{d\sigma }{d\Omega }_e }
\right)_{Mott} $ is the scattering cross-section on a spinless
structureless particle obtained by Mott; $p$ is the transmitted
deuteron momentum in units fm-1; $A(p)$ and $B(p)$ are the
electric and magnetic structure functions (or structure functions
determined by the electromagnetic structure of a deuteron)

\begin{equation}
\label{eq2}
A = G_C^2 + \frac{8}{9}\eta ^2G_Q^2 + \frac{2}{3}\eta G_M^2 ;
\end{equation}

\begin{equation}
\label{eq3}
B = \frac{4}{3}\eta \left( {1 + \eta } \right)G_M^2 ,
\end{equation}

where $\eta = \frac{p^2}{4m_d^2 }$; $m_{d}$ -- is the mass of
deuteron. The charge $G_{C}(p)$, quadrupole $G_{Q}(p)$ and
magnetic $G_{M}(p)$ deuteron form factors (FFs) contain
information about the electromagnetic characteristics of a
deuteron \cite{Gourdin1963, Gross1964, Gari1976, Mcgurk1977,
Lomon1980, Gilman2002, Adamuscin2008}:

\begin{equation}
\label{eq4}
G_C = G_{EN} D_C ;
\quad
G_Q = G_{EN} D_Q ;
\quad
G_M = \frac{m_d }{2m_p }\left( {G_{MN} D_M + G_{EN} D_E } \right).
\end{equation}

The body form factors $D_{i}$ are determined by DWFs in the coordinate
representation

monopoly electric $D_C = \int\limits_0^\infty {\left[ {u^2 + w^2}
\right]j_0 dr} $;

quadrupole electric $D_Q = \frac{3}{\sqrt 2 \eta
}\int\limits_0^\infty {\left[ {uw - \frac{w^2}{\sqrt 8 }}
\right]j_2 dr} $;

transverse magnetic $D_M = 2\int\limits_0^\infty {\left[ {\left(
{u^2 - \frac{w^2}{2}} \right)j_0 + \left( {\frac{uw}{\sqrt 2 } +
\frac{w^2}{2}} \right)j_2 } \right]dr} $;

longitudinal magnetic $D_E = \frac{3}{2}\int\limits_0^\infty
{w^2\left[ {j_0 + j_2 } \right]dr} $;

where $G_{EN} = G_{Ep} + G_{En} $; $G_{MN} = G_{Mp} + G_{Mn} $ are
the isoscalar electric and magnetic FFs; $G_{Ep}$ and $G_{En}$ are
proton and neutron isoscalar electric FFs; $G_{Mp}$ and $G_{Mn}$
are proton and neutron isoscalar magnetic FFs; $j_{0}$, $j_{2}$
are spherical Bessel functions of zero and second order from
argument \textit{pr}/2.

The original dipole fit (DFF) is the simplest representation for
the proton and neutron FFs \cite{Haftel1980, Bekzhanov2013}:

\begin{equation}
\label{eq5}
G_{Ep} = F_N ;
\quad
G_{En} = 0;
\quad
G_{Mp} = \mu _p G_{Ep} ;
\quad
G_{Mn} = \mu _n G_{Ep} ;
\end{equation}

Here the nucleon FF is written in the form of a dipole
\cite{Haftel1980}

\begin{equation}
\label{eq6}
F_N (p^2) = \left( {1 + \frac{p^2}{0.71(GeV / c)^2}} \right)^{ - 2} = \left(
{1 + \frac{p^2}{18.235fm^{ - 2}}} \right)^{ - 2}.
\end{equation}

Analysing formulas (\ref{eq1})-(\ref{eq3}), it will be obvious
\cite{Krutov2002} that the angular dependence of the differential
cross-section allows us to independently measure the structure
functions $A$ and $B$. However, experiments for unpolarized
particles do not give complete information about all quantities
describing ed-scattering and not allow dividing the contributions
of charge and quadrupole FFs in to $A$. Therefore, additional
experiments with polarized particles are required for a complete
description: 1) scattering on a polarized deuteron target; 2)
measurement of tensor polarization of recoil deuterons.

Experimental data for the electric structure function $A(p)$ are
given in the papers of Stanford \cite{Buchanan1965}, Orsay
\cite{Benaksas1966, Grossetete1966}, CEA \cite{Elias1969}, DESY
\cite{Galster1971}, SLAC \cite{Arnold1975, Martin1977}, Mainz
\cite{Simon1981}, Bonn \cite{Cramer1985}, Saclay
\cite{Platchkov1990}, Bates \cite{Garcon1994}, JLab
\cite{Abbott1999, Alexa1999}, JLab2007 \cite{PR08019}
collaborations and in Garcon review \cite{Garcon1994}.

\textbf{3. Asymptotic of the electric structure function}

In quantum chromodynamics (QCD) \cite{Brodsky1992, Gutsche2016},
with large momentums, asymptotic values and their relations for
the structure functions and for deuteron FFs are written as:

\begin{equation}
\label{eq7}
\sqrt A \sim \sqrt B \sim G_C \sim \frac{1}{p^{10}};
\quad
G_Q \sim G_M \sim \frac{1}{p^{12}};
\quad
B:A:G_C^2 = 4:1:\frac{1}{3};
\end{equation}

\begin{equation}
\label{eq8}
G_C :G_M :G_Q = \left( {1 - \frac{2}{3}\eta } \right):2: - 1.
\end{equation}

In addition to the representation of the structure functions $A$
and $B$, in paper \cite{Salme2000} there is a form of presentation
of the results in the form $A(p^2) / (F_N^2 F)$ and $B(p^2) /
(F_N^2 F_1 )$, and also according to \cite{Lev2000} in the form as
$\Gamma _M (p^2) / (F_N^2 F_1 )$, where $\Gamma _M (p^2) = \left[
{G_M (p^2)m_p / (\mu _d m_d )} \right]^2$; $F = (1 + p^2 / 0.1)^{
- 2.5}$; $F_1 = (1 + p^2 / 0.1)^{-3}$. Values $F$ and $F_1$ in
dimensions [(GeV/c)$^{2}$] for $p^{2}$. Determination of structure
function $A(p)$ is important for the study of the deuteron FF

\begin{equation}
\label{eq9}
F_d (p^2) = \sqrt {A(p^2)}
\end{equation}

and of the reduced deuteron FF \cite{Brodsky1976}

\begin{equation}
\label{eq10}
f_d (p^2) = F_d (p^2) / F_N^2 \left( {\frac{p^2}{4}} \right).
\end{equation}

The deuteron FF in QCD is defined as \cite{Brodsky1983}

\begin{equation}
\label{eq11}
F_d (p^2) \approx \left[ {\frac{\alpha _S (p^2)}{p^2}}
\right]^5\sum\limits_{m,n} {d_{mn} \left[ {\ln \left( {\frac{p^2}{\Lambda
^2}} \right)} \right]^{ - \gamma _n^d - \gamma _m^d }} \left[ {1 + O\left(
{\alpha _S (p^2);\frac{m}{p}} \right)} \right].
\end{equation}

In the predictions of QCD \cite{Brodsky1983}, the reduced FF at
$p^{2}$>2 (GeV/c)$^{2}$ has the following asymptotic

\begin{equation}
\label{eq12}
f_d (p^2) \approx \frac{\alpha _S (p^2)}{p^2}\left[ {\ln \left(
{\frac{p^2}{\Lambda ^2}} \right)} \right]^{ - \lambda },
\end{equation}

where $\lambda = \frac{2}{5}\frac{C_F }{\beta } = -
\frac{8}{145}$; $C_F = \frac{N_c ^2 - 1}{2N_c }$; $\beta = 11 -
\frac{2}{3}N_f $. Here $N_{c}$=3 and $N_{f}$=2 are the numbers of
colours and flavors of quarks.

It was shown in \cite{Brodsky1983} that the structure function
(\ref{eq2}) can be decomposed by the nucleon FF $F_{N}(p^{2}$/4) and
the so-called "reduced" nuclear FF $f_{d}(p^{2})$:

\begin{equation}
\label{eq13}
\sqrt {A(p^2)} = f_d (p^2)F_N^2 (p^2 / 4).
\end{equation}

Parameterization was used to determine the reduced deuteron FF as
\cite{Rekalo2003}

\begin{equation}
\label{eq14}
f_d (p^2) = N\frac{\alpha _S (p^2)}{p^2}\left[ {\ln \left(
{\frac{p^2}{\Lambda ^2}} \right)} \right]^{ - \Gamma },
\end{equation}

where $N$ is the normalization factor; $\alpha _S (p^2) = \left[
{\ln \left( {\frac{p^2}{\Lambda ^2}} \right)} \right]^{-1}$ are
strong interaction coupling constant; $\Lambda $ is the QCD-scale
parameter; $\Gamma = - 8 / 145$ is determined by the leading
anomalous dimension.

The asymptotic for the deuteron FF at large momentums are
determined by the expression \cite{Krutov2004}:

\begin{equation}
\label{eq15}
F_d (p^2) \sim \frac{\mbox{1}}{\left( {p^2} \right)^m},
\end{equation}

where $m$=7/2 and 13/4 are for non-relativistic and relativistic
impulse approximations (NRIA and RIA) respectively. In addition,
in paper \cite{Krutov2005} it is noted that the magnetic FF
$G_{M}$ makes a major contribution to the structure function
$A(p)$ and, accordingly, to the deuteron FF $F_{d}$ (in the
non-relativistic IA):

\begin{equation}
\label{eq16}
A^{(NR)}(p^2\mbox{)} \sim \frac{1}{\left( {p^2} \right)^6}\frac{(\mu _p +
\mu _n )^2(m_0^2 )^4}{3\pi ^2m_d^2 }\frac{\left[ {\sum\limits_j {C_j m_j^2 }
} \right]^2\left[ {\sum\limits_j {C_j / m_j^2 } } \right]^4}{\left[
{\sum\limits_j {C_j / m_j^4 } } \right]^2};
\quad
F_d^{(NR)} (p^2) \sim \frac{\mbox{1}}{\left( {p^2} \right)^3};
\end{equation}

where $m_0^2 = 0.71(GeV / c)^2$; $C_{j}$, $m_{j}$ are the
expansion coefficients of the S-component of DWF in the momentum
representation. In the RIA, the asymptotics for $A$ and $F_{d}$
are defined as follows \cite{Krutov2005}:

\begin{equation}
\label{eq17}
A^{(R)} \sim \frac{m_d }{p}A^{(NR)};
\quad
F_d^{(R)} \sim \frac{F_d^{(NR)} }{\sqrt p }.
\end{equation}

In addition, in \cite{Krutov2006} takes into account the influence
of the ``non-Rosenbluth'' behaviour of proton FFs on deuteron FFs
and, as a consequence, the major contribution to the asymptotic of
the structure function $A(p)$ gives the magnetic deuteron FF and
the ``non-Rosenbluth'' behaviour of proton FFs enhances this
effect. The obtained asymptotics \cite{Krutov2006} for $A$ and
$F_{d}$ in NRIA (NR) and RIA (R) are written as

\begin{equation}
\label{eq18}
A^{(NR)}(p^2\mbox{)} \sim \frac{1}{\left( {p^2} \right)^8}\frac{2048(\mu _p
+ \mu _n )^2(m_0^2 )^4}{3\pi m_d^2 }\left[ {\sum\limits_j {C_j m_j^2 } }
\right]^4;
\quad
F_d^{(NR)} (p^2) \sim \frac{\mbox{1}}{\left( {p^2} \right)^4};
\end{equation}

\begin{equation}
\label{eq19}
A^{(R)} \sim \frac{p^3}{32m_d^3 }A^{(NR)};
\quad
F_d^{(R)} \sim p^{3 / 2}F_d^{(NR)} .
\end{equation}

As stated in \cite{Krutov2006}, given the experimental data JLab
\cite{Abbott1999, Alexa1999} for structure function $A(p)$ for
momentum values of $p^{2}\sim $6~(GeV/c)$^{2}$, the deuteron FF
can be interpolated by the function

\begin{equation}
\label{eq20}
F_d^{(\exp )} \sim \frac{1}{\left( {p^2} \right)^{3.76\pm 0.41}}.
\end{equation}

The perturbative QCD (pQCD) is used to calculate the rate of fall
of deuteron FFs for large values $p^{2}$ \cite{VanOrden2005},
where for a structure function (\ref{eq2}) is valid the
approximation $A(p^2) \sim c_0 / p^{2n}$ at $n$=9; 10.

The deuteron FF was obtained in the high-quality description of
the deuteron electromagnetic FFs in a soft-wall anti-de
Sitter/quantum chromodynamics approach (AdS/QCD) as
\cite{Gutsche2015}

\begin{equation}
\label{eq21}
F_d(p^{2}) \equiv F(p^{2})=
\frac{\Gamma(6)\Gamma(a+1)}{\Gamma(a+6)},
\end{equation}

where $F$ is the twist-6 hadronic form factor;
$a^{2}=p^{2}$/(4$\kappa ^{2})$; the scale parameter $\kappa $ in
the range of 150 MeV<$\kappa $<250 MeV.

\textbf{4. Calculations and conclusions}

One of the simplest and most convenient analytical forms of DWFs
in coordinate representation is the following \cite{arxiv05174}

\begin{equation}
\label{eq22}
\left\{ {\begin{array}{l}
 u(r) = r\sum\limits_{i = 1}^N {A_i e^{ - a_i r},} \\
 w(r) = r^3\sum\limits_{i = 1}^N {B_i e^{ - b_i r}.} \\
 \end{array}} \right.
\end{equation}

The number of expansions in (\ref{eq22}) for Reid93 potential was $N$=29 since in this
case, the calculated deuteron parameters by these forms will be good
(deuteron radius $r_{m}$, electric quadrupole moment, magnetic moment, the
D-state probability, asymptotic D/S-states, etc.).

In addition to the simplest DFF according to (\ref{eq5}), the
calculations of the structure function $A(p)$ for other
theoretical models and the approximations for the nucleon form
factors were performed. The first group includes the following:
the modified dipole fit 2 (labelled as MDFF2)
\cite{Bekzhanov2013}, the relativistic harmonic oscillator model
(RHOM) based on the quark model with the relativistic oscillator
potential \cite{Bekzhanov2014, Burov1993}, the vector meson
dominance model (VMDM) \cite{Iachello1973} taking into account the
parameters of the full Gari-Krumpelmann model \cite{Gari1985}; and
the second - Kelly parameterization (Kelly) \cite{Kelly2004}, the
approximation based JLab measurements of the isoscalar electric
and magnetic FFs (JLab) \cite{Gilman2002} and Bradford
parameterization (Bradford) \cite{Bradford2006}.

Fig. 1 shows the results of the calculations of the electric
structure function $A(p)$ by selecting seven specified sets of
nucleon form FFs and applying of DWF (\ref{eq22}) for the
nucleon-nucleon potential Reid93. Theoretical calculations in
comparison with experimental data \cite{Buchanan1965,
Benaksas1966, Grossetete1966, Elias1969, Galster1971, Arnold1975,
Martin1977, Simon1981, Cramer1985, Platchkov1990, Garcon1994,
Abbott1999, Alexa1999, PR08019, Garcon1994}. The best agreement of
the theoretical results with the experiment is available for
calculations when choosing VMDM for momentums up $p$=13 fm$^{ -
1}$, and for approximations, there will be coincidence only for
momentum at $p \le $5 fm$^{-1}$.

\pdfximage width 150mm {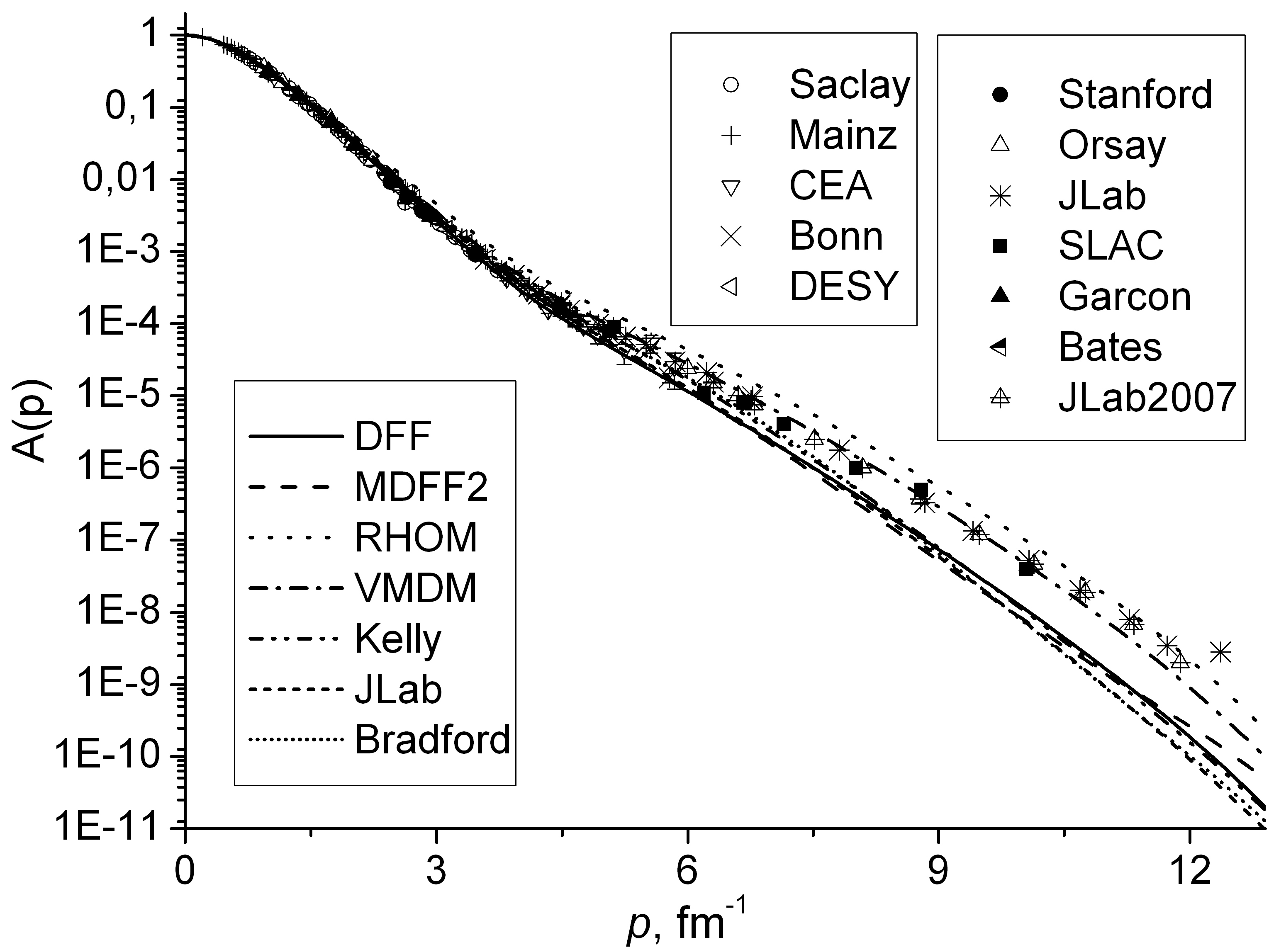}\pdfrefximage\pdflastximage

Fig. 1. The electric structure function $A(p)$. The theoretical
calculations for DWF (\ref{eq22}) with models \cite{Haftel1980,
Bekzhanov2013, Bekzhanov2014, Burov1993, Iachello1973, Gari1985,
Kelly2004, Gilman2002, Bradford2006} are compared with the
experimental data from \cite{Buchanan1965, Benaksas1966,
Grossetete1966, Elias1969, Galster1971, Arnold1975, Martin1977,
Simon1981, Cramer1985, Platchkov1990, Garcon1994, Abbott1999,
Alexa1999, PR08019, Garcon1994}

In \cite{VLvNU56} for DWFs (\ref{eq22}) obtained the deuteron FFs
$G_{i}$, which is determined by the coefficients of the analytical
forms for DWFs, nucleon isoscalar FFs and the order of momentum

\begin{equation}
\label{eq23}
G_C \approx 32G_{EN} \sum\limits_{i,j = 1}^N {\frac{A_i A_j a_{ij} }{\left(
{p^2 + 4a_{ij}^2 } \right)^2}} ;
\quad
G_Q \approx - \frac{9216G_{EN} }{\sqrt 2 \eta }\sum\limits_{i = 1}^N
{\frac{A_i B_i (a_i + b_i )p^2}{\left( {p^2 + 4(a_i + b_i )^2} \right)^4}}
;
\end{equation}

\begin{equation}
\label{eq24}
G_M \approx \frac{m_d }{m_p }G_{MN} \left( {32\sum\limits_{i,j = 1}^N
{\frac{A_i A_j a_{ij} }{\left( {p^2 + 4a_{ij}^2 } \right)^2}} -
\frac{3072}{\sqrt 2 }\sum\limits_{i = 1}^N {\frac{A_i B_i (a_i + b_i
)p^2}{\left( {p^2 + 4(a_i + b_i )^2} \right)^4}} } \right).
\end{equation}

where $a_{ij} = a_i + a_j $; $b_{ij} = b_i + b_j $. And,
substituting in (\ref{eq23}) and (\ref{eq24}) the values of $G_{EN}$,
$G_{MN}$ (for DFF) and $\eta $, we can write the asymptotics of
the deuteron FFs for large momentum values in the form
\cite{VLvNU56}:

\begin{equation}
\label{eq25}
G_C \sim \frac{1}{p^8};
\quad
G_Q \sim \frac{1}{p^{12}};
\quad
G_M \sim \frac{1}{p^8}.
\end{equation}

In formulas (\ref{eq25}) takes into account  only the leading
parts at large momentum and the simplification of records without
the expansion coefficients of DWF (\ref{eq22}). When using
asymptotics of deuteron FFs at large momentums (\ref{eq25}), we
find the asymptotic of the electric structure function (\ref{eq2})
as $A \sim \left( {\frac{1}{p^8}} \right)^2 + \eta ^2\left(
{\frac{1}{p^{12}}} \right)^2 + \eta \left( {\frac{1}{p^8}}
\right)^2$, which will be determined by the third part, that is,
the magnetic FF:

\begin{equation}
\label{eq26}
A \sim \frac{1}{p^{14}}.
\end{equation}

The exact value of asymptotic for $A(p)$ can be obtained by considering
expressions (\ref{eq23}) and (\ref{eq24}).

Comparison of our asymptotic (\ref{eq26}) for DWFs (\ref{eq22})
with other asymptotics (\ref{eq15})-(\ref{eq19}) in NRIA and RIA
\cite{Krutov2005, Krutov2006} for which $A \sim \frac{1}{p^k}$ (at
$k$=12; 13; 16; 13 respectively) indicates their similarity.

According to the approximation \cite{Krutov2006} according to
formula (\ref{eq20}) for the electric structure function $A(p)$ is
a fair expression $\sqrt {A^{(approx1)}} \sim \frac{1}{\left(
{p^2} \right)^{3.76}}$. The approximation of the experimental data
of the JLab \cite{Alexa1999} and JLab2007 \cite{PR08019}
collaborations in the momentum interval at $p \approx $10.6-12.4
fm$^{-1}$ using the function $\frac{1}{\left( {p^2} \right)^N}$
similar to (\ref{eq15}) gives the result for $N \approx $3.795575,
i.e. $\sqrt {A^{(approx2)}} \sim \frac{1}{\left( {p^2}
\right)^{3.795}}$. In Fig. 2, the results of these two
approximations are labelled as approx1 and approx2 respectively,
and the calculations using the DWF (\ref{eq22}) for Reid93
potential at different nucleon FFs are indicated as lines.

The approximation of the calculated values $A(p)$ in the range momentums at
$p$=10.6-12.5 fm$^{ - 1}$ using the function $p^{ - N}$ showed, that the value
$N$ was obtained as 8.44, 8.50, 7.48, 7.82, 8.50, 8.68 and 8.69 for DFF, MDFF2,
RHOM, VMDM, JLab, and Bradford respectively. That is, closer to the
experimental approximation are the values for RHOM and VMDM.

To estimate the quality of the approximation, the value
\textit{$\chi $}$^{2}$ is calculated, which per degree of freedom
of function. Compared to the experiment, the value \textit{$\chi
$}$^{2}$ (multiplied by a factor 10$^{-16})$ for approx1,
approx2, RHOM, VMDM is 1.21, 1.63, 3.46, 0.27 respectively.
Obviously, the ``best'' approximation is approx1, and of the two
variants the RHOM model is ``worse''.

\pdfximage width 150mm {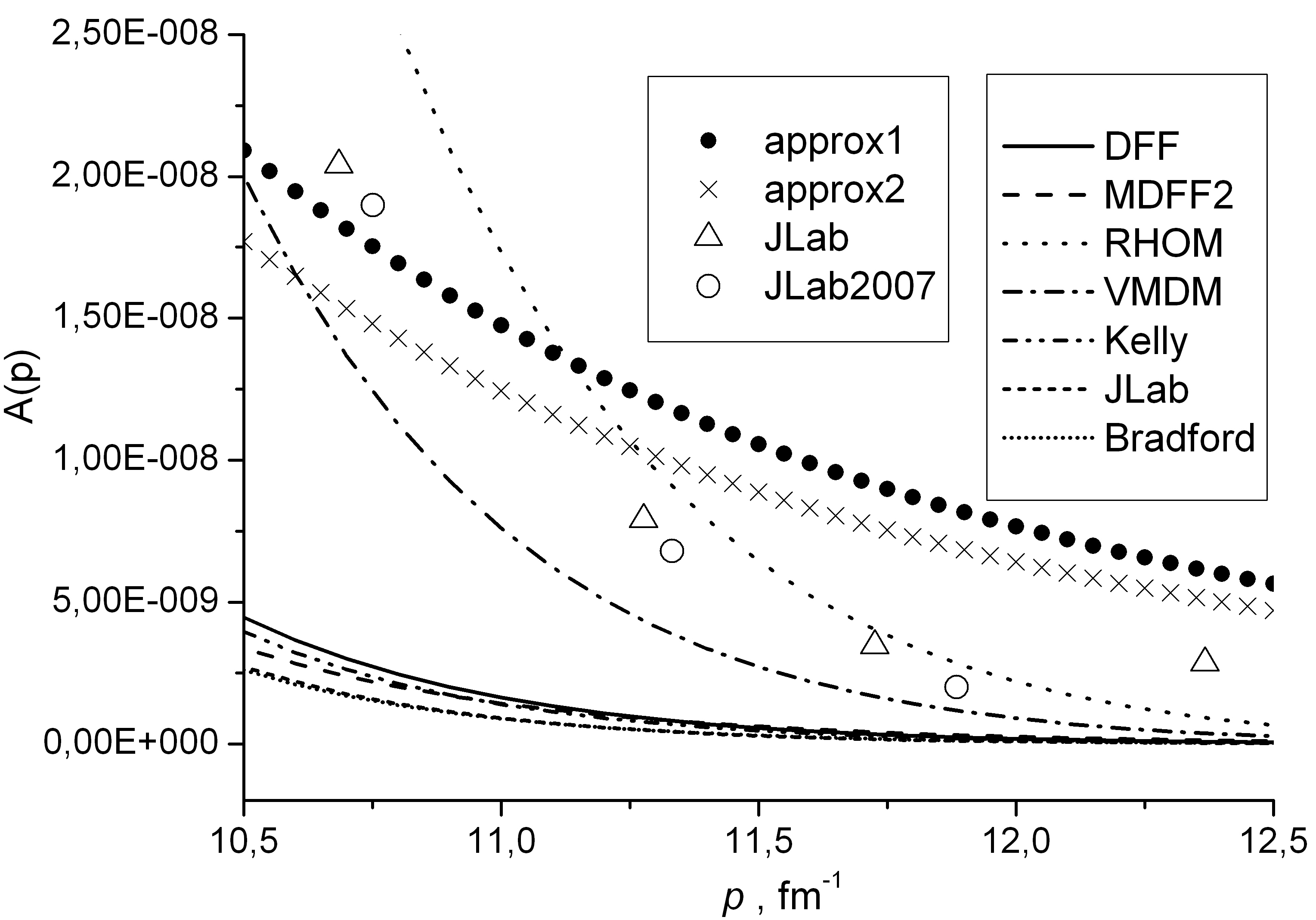}\pdfrefximage\pdflastximage

Fig. 2. The electric structure function $A(p)$ at large values of
the momentum. The theoretical calculations for DWF (\ref{eq22})
with models \cite{Haftel1980, Bekzhanov2013, Bekzhanov2014,
Burov1993, Iachello1973, Gari1985, Kelly2004, Gilman2002,
Bradford2006} are compared with the experimental data from
\cite{Alexa1999, PR08019} and with the two approximations

As can be seen from Figs. 1 and 2 almost all (except for the
results for RHOM) the theoretical values obtained for the electric
structure function $A(p)$ are below the experimental data. Similar
behaviour of $A(p)$ was obtained in \cite{Krutov2005, Krutov2006}
for NRIA and RIA, for Moscow, NijmI, NijmII, CD-Bonn, Paris
potentials in \cite{Khokhlov2015}, for calculations in chiral
effective field theory \cite{Piarulli2013} at leading order (LO)
and with inclusion of charge operators up to N$^{3}$LO, for NRIA
\cite{Adamuscin2008} predictions using non-dipole and dipole
behaviour $G_{Ep}(p)$, for phenomenological potentials in NRIA
(Bonn-A, B, C, Q, Reid-SC, Paris and Argonne v18 at $p$>4 fm$^{-
1}$ \cite{Garcon2001}), for generalized IA \cite{Huang2009} at
$p$>5 fm$^{-1}$, for equal time approximation and relativistic
quasipotential approximation of the
Blankenbecler-Sugar-Logunov-Tavkhelidze \cite{Hummel1994}.

Below the experimental data are the calculations $A(p)$
\cite{Allen2001} (at $p$>3.6 fm$^{-1})$ for Reid93 potential using
the Gari-Krumpelmann and Mergell-Meissner-Drechsel
parameterizations of the nucleon FFs.

As the analysis shows, above the experimental data are the values
$A(p)$, that were calculated for OBEPQ-A, B, C potentials
\cite{Arenhovel2000} (for momentum values at $p$>3 fm$^{-1})$, for
IA with the $\rho \pi \gamma $ MEC included \cite{Phillips20002}
(at $p$>9 fm$^{-1})$, for Bonn (Bonn-B, FULLF, OBEPF) and Nijmegen
(Nijm93, NijmI, NijmII) potentials of the Gari-Krumpelmann nucleon
FFs \cite{Plessas1995} (at $p$>4 fm$^{-1})$.

For the momentum interval at 7.2-13 fm$^{-1}$, two approximations
for $A(p)$ were performed in \cite{Garcon2001}. The first
approximation is obtained for dimensional scaling in pQCD
\cite{Brodsky1975}, wherein the case of a hadron composed of $n$
quarks, one can expect that the leading FF has asymptotic as $p^{
- 2(n - 1)}$. Since this true in both the initial and final
states, so the FF of the system is written as $F^2 \sim p^{ - 4(n
- 1)}$, and for deuteron ($n$=6) the asymptotic for the structure
function will be as $A \sim p^{-20}$. The second approximation in
\cite{Garcon2001} is as $A \sim p^{-16}$ follows. The fit of the
five data for the highest $p^{2}$ from \cite{Alexa1999} using $A =
p^{ - 2m}$ gives the value $m$=8.0. If we exclude the inaccurate
last point, then $m$=8.7. The experimental data point
\cite{Alexa1999} at 5.955 (GeV/c)$^{2}$ (or 12.37 fm$^{-1})$ does
not lie on the approximation dependence $A = p^{ - 2m}$ at
$m$=8.0. All curves in paper \cite{Garcon2001} (see Fig. 5.17) are
normalized to the point at $p^{2}$=11.275 fm$^{-1}$.

Fig. 3 shows a set of approximations. Similar to
\cite{Garcon2001}, these approximations are normalized about the
point $p$=11.276 fm$^{-1}$. In Fig. 3, the labelled ``QCD''
corresponds to the application of formula (\ref{eq14}). The
parameters for (\ref{eq14}) are selected from paper
\cite{Rekalo2003} for the case $F_N^2 \to G_{Ep}^2 $ as follows:
$N$=0.16; $\Lambda $=0.20 GeV. Here the proton electric FF was
written as (values 0.129$p^{2}$ in dimension [(GeV/c)$^{2}$])

$G_{Ep} = \frac{1}{(1 + p^2 / m_D^2 )^2}(1 - 0.129p^2)$at $m_D^2 = 0.71GeV^2$.

In Fig. 3 uses the following designations for approximations: NRIA
\cite{Krutov2006} as $A \sim p^{-16}$; approximation for pQCD
\cite{VanOrden2005, Garcon2001} as $p^{-20}$; analytical result
(\ref{eq26}) for DWFs (\ref{eq22}) with DFF in form $p^{-14}$;
NRIA \cite{Krutov2005} as $p^{-12}$; ``Approx2'' described above;
the approximation $A = p^{ - 2m}$ at $m$=8.7 \cite{Garcon2001} is
labelled as $p^{-17.4}$; the designation ``Global fit''
corresponds to the approximation of the experimental data for the
momentum interval $p$=4.00874-12.36678 fm$^{-1}$ with the
resultant function $\sqrt A \sim p^{-5.787}$.

\pdfximage width 150mm {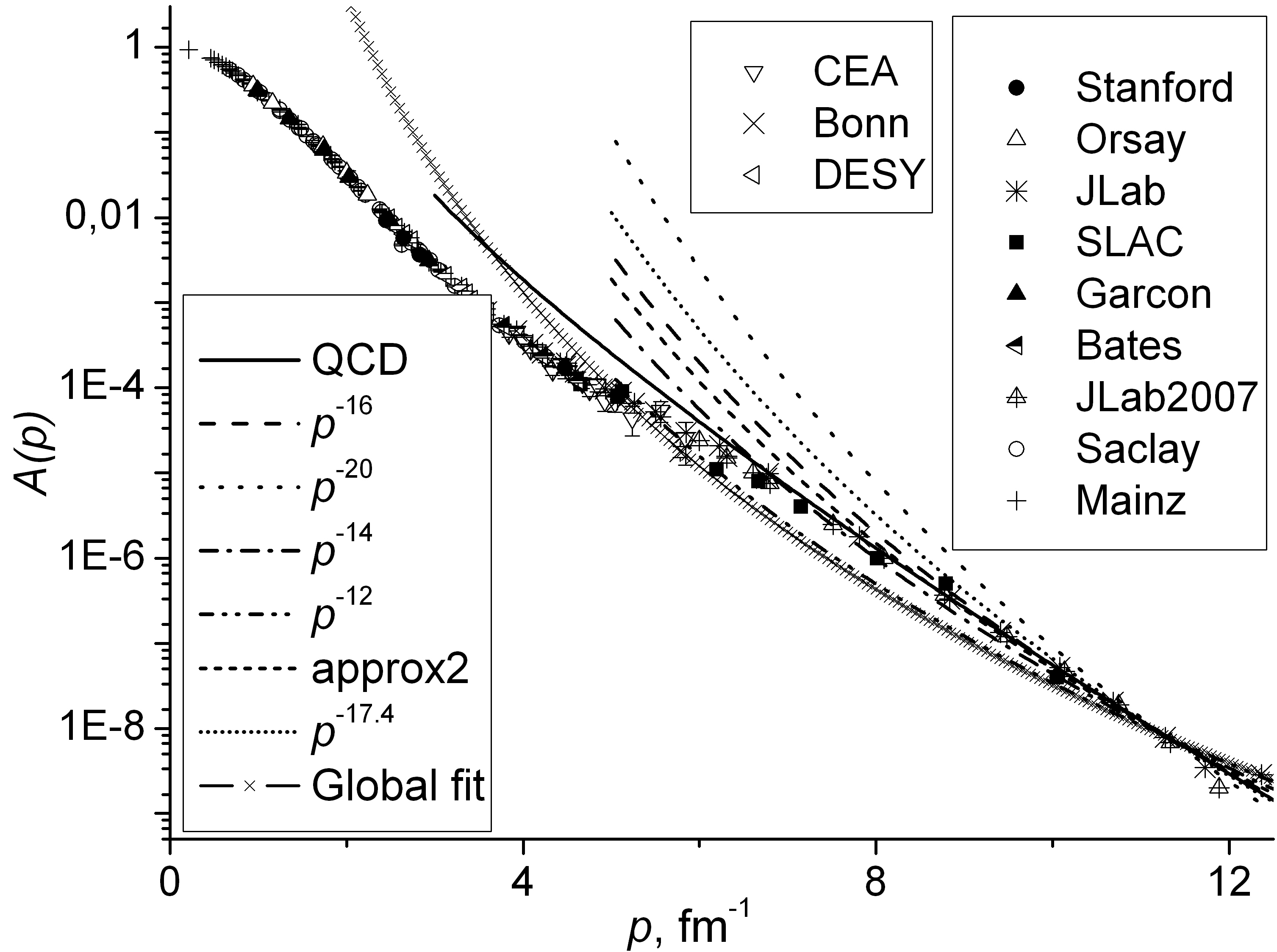}\pdfrefximage\pdflastximage

Fig. 3. Approximations of the electric structure function $A(p)$.
Approximations and global fit are compared with the experimental
data from \cite{Buchanan1965, Benaksas1966, Grossetete1966,
Elias1969, Galster1971, Arnold1975, Martin1977, Simon1981,
Cramer1985, Platchkov1990, Garcon1994, Abbott1999, Alexa1999,
PR08019, Garcon1994}

So, we can draw the following conclusions:

1. Theoretical features of the estimation of the asymptotic
behaviour of the electric structure function $A(p)$ at large
momentums are analysed. Asymptotic values of $A(p)$ were obtained
for Reid93 potential when using different models and
approximations for nucleon FFs. The approximation of the values
$A(p)$ in the range momentums at $p$=10.6-12.5 fm$^{-1}$ showed,
that closer to the experimental approximation are the values for
RHOM and VMDM.

2. For electric structure function, the basic theoretical forms of
asymptotic behaviour and approximations according to the
experimental data and in comparison with the experiment of the
leading collaborations are demonstrated. In addition to the QCD
approach, the asymptotic for $A(p)$ is represented in the form of
the power function $p^{-n}$.

3. The asymptotic behaviour of $A(p)$ (\ref{eq26}) has been determined by taking into
account the asymptotic behaviour of deuteron FFs (\ref{eq25}) and DFF for nucleon
FFs. The exact value of $A(p)$ at large momentum values is determined by the
asymptotics of the deuteron FFs $G_{i}$, the coefficients of the analytical
forms of the DWs and the nucleon isoscalar FFs.

4. The following numerical calculations of the ratio $B/A$ for
momentums up $p$=12 fm$^{-1}$ \cite{Brodsky1992, Holt2012}, the
tensor $t_{2j}$ and vector $t_{1i}$ polarizations
\cite{Gilman2002}, the spin correlation coefficients and tensor
asymmetries \cite{Gakh2014} and other polarization observables
taking into account the behaviour of the electric structure
function are promising.


\begin{thebibliography}{0}


\bibitem{Hasell2011} D.K. Hasell et al., Annu. Rev. Nucl. Part. Sci. \textbf{61},
409 (2011).

\bibitem{Alexa1999} L.C. Alexa et al., Phys. Rev. Lett. \textbf{82, }1374 (1999).

\bibitem{Abbott1999} D. Abbott et al., Phys. Rev. Lett. \textbf{82, }1379 (1999).

\bibitem{PR08019} R. Alarcon et al., Jefferson Lab PAC33 Proposal December 2007
PR-08-019.

\bibitem{Gilman2002} R. Gilman, F. Gross, J. Phys. G. \textbf{28}, R37 (2002).

\bibitem{Gakh2014} G.I. Gakh, A.G. Gakh, E. Tomasi-Gustafsson, Phys. Rev. C
\textbf{90}, 064901 (2014).

\bibitem{Phillips2005} D.R. Phillips, S.J. Wallace, N.K. Devine, Phys. Rev. C
\textbf{72}, 014006 (2005).

\bibitem{Kohl2008} M. Kohl, Nucl. Phys. A \textbf{805}, 361 (2008).

\bibitem{Piarulli2013} M. Piarulli et al., Phys. Rev. C \textbf{87}, 014006 (2013).

\bibitem{Khokhlov2015} N.A. Khokhlov, A.A. Vakulyuk, Phys. Atom. Nucl., \textbf{78},
92 (2015).

\bibitem{Gutsche2016} T. Gutsche, V.E. Lyubovitskij, I. Schmidt, Phys. Rev. D
\textbf{94}, 116006 (2016).

\bibitem{Stoks1994} V.G.J. Stoks et al., Phys. Rev. C \textbf{49}, 2950 (1994).

\bibitem{Swart1995} J.J. de Swart et al., Few-Body Syst. Suppl. \textbf{8}, 438
(1995).

\bibitem{MPLA3125} V. I. Zhaba, Mod. Phys. Lett. A \textbf{31}, 1650139 (2016).

\bibitem{Abbott20001} D. Abbott et al., Phys. Rev. Lett. \textbf{84}, 5053 (2000).

\bibitem{Elias1969} J.E. Elias et al., Phys. Rev. \textbf{177, }2075 (1969).

\bibitem{Galster1971} S. Galster et al., Nucl. Phys. B \textbf{32, }221 (1971).

\bibitem{Donnelly1986} T.W. Donnelly, A.S. Raskin, Ann. Phys. (N. Y.) \textbf{169},
247 (1986).

\bibitem{Garcon2001} M. Garcon, J.W. van Orden, Adv. Nucl. Phys. \textbf{26}, 293
(2001).

\bibitem{Rosenbluth1950} M.N. Rosenbluth, Phys. Rev. \textbf{79}, 615 (1950).

\bibitem{Gourdin1963} M. Gourdin, Nuovo Cimento, \textbf{28}, 533 (1963).

\bibitem{Gross1964} F. Gross, Phys. Rev. \textbf{136}, B140 (1964).

\bibitem{Gari1976} M. Gari, H. Hyuga, Nucl. Phys. A \textbf{264}, 409 (1976).

\bibitem{Mcgurk1977} N.J. McGurk, H. Fiedeldey, Nucl. Phys. A, \textbf{281},
310(1977).

\bibitem{Lomon1980} E.L. Lomon, Ann. Phys. \textbf{125}, 309 (1980).

\bibitem{Adamuscin2008} C. Adamusc\'{\i}n et al., Phys. Rev. C \textbf{78}, 025202
(2008).

\bibitem{Haftel1980} M.I. Haftel, L. Mathelitsch, H.F.K. Zingl, Phys. Rev. C
\textbf{22}, 1285 (1980).

\bibitem{Bekzhanov2013} A. Bekzhanov et al., Nucl. Phys. B (Proc. Suppl.)
\textbf{245}, 65 (2013).

\bibitem{Krutov2002} A.F. Krutov, Theor. Phys. \textbf{3}, 5 (2002).

\bibitem{Buchanan1965} C.D. Buchanan, M.R. Yearian, Phys. Rev. Lett. \textbf{15,
}303 (1965).

\bibitem{Benaksas1966} D. Benaksas et al., Phys. Rev. \textbf{148, }1327 (1966).

\bibitem{Grossetete1966} B. Grossetete et al., Phys. Rev. \textbf{141, }1425 (1966).

\bibitem{Arnold1975} R. Arnold et al., Phys. Rev. Lett. \textbf{35, }776 (1975).

\bibitem{Martin1977} F. Martin et al., Phys. Rev. Lett. \textbf{38}, 1320 (1977).

\bibitem{Simon1981} G.G. Simon et al., Nucl. Phys. A \textbf{364, }285 (1981).

\bibitem{Cramer1985} R. Cramer, M. Renkhoff, J. Drees et al., Z. Phys. C
\textbf{29}, 513 (1985).

\bibitem{Platchkov1990} S. Platchkov et al., Nucl. Phys. A \textbf{510}, 740 (1990).

\bibitem{Garcon1994} M. Garcon et al., Phys. Rev. C \textbf{49}, 2516 (1994).

\bibitem{Brodsky1992} S.J. Brodsky, J.R. Hiller, Phys. Rev. D \textbf{46}, 2141
(1992).

\bibitem{Salme2000} G. Salme, F. M. Lev, E. Pace, Few-Body Syst. Suppl. \textbf{12},
235 (2000).

\bibitem{Lev2000} F.M. Lev, E. Pace, G. Salme, Phys. Rev. C \textbf{62}, 064004
(2000).

\bibitem{Brodsky1976} S.J. Brodsky, B.T. Chertok, Phys. Rev. D \textbf{14}, 3003
(1976).

\bibitem{Brodsky1983} S.J. Brodsky, C.-R. Ji, G.P. Lepage, Phys. Rev. Lett.
\textbf{51}, 83 (1983).

\bibitem{Rekalo2003} M.P. Rekalo, E. Tomasi-Gustafsson, Eur. Phys. J. A \textbf{16},
563 (2003).

\bibitem{Krutov2004} A.F. Krutov, V.E. Troitsky, N.A. Tsirova, Theor. Phys.\textbf{
5}, 17 (2004).

\bibitem{Krutov2005} A.F. Krutov, V.E. Troitsky, N.A. Tsirova, Theor. Phys.\textbf{
6}, 71 (2005).

\bibitem{Krutov2006} A.F. Krutov, V.E. Troitsky, N.A. Tsirova, Vestnik SamSU,
Natural Scie. Ser. \textbf{3(43)}, 100 (2006).

\bibitem{VanOrden2005} J.W. van Orden. Electron Scattering (Kluwer Academic/Plenum
Publishers, New York, 2005) 279-289 p.

\bibitem{Gutsche2015} T. Gutsche et al., Phys. Rev. D \textbf{91}, 114001 (2015).

\bibitem{arxiv05174} V.I. Zhaba, e-print arXiv:nucl-th/1603.05174

\bibitem{Bekzhanov2014} A.V. Bekzhanov, S.G. Bondarenko, V.V. Burov, JETP Lett.
\textbf{99}, 613 (2014).

\bibitem{Burov1993} V.V. Burov et al., Europhys. Lett. \textbf{24}, 443 (1993).

\bibitem{Iachello1973} F. Iachello, A.D. Jackson, A. Lande, Phys. Lett.
\textbf{43B}, 191 (1973).

\bibitem{Gari1985} M. Gari, W. Krumpelmann, Z. Phys. A \textbf{322}, 689 (1985).

\bibitem{Kelly2004} J.J. Kelly, Phys. Rev. C \textbf{70}, 068202 (2004)

\bibitem{Bradford2006} R. Bradford et al., Nucl. Phys. B (Proc. Suppl.)
\textbf{159}, 127 (2006)

\bibitem{VLvNU56} V.I. Zhaba, Visnyk Lviv Univ. Ser. Phys. \textbf{56}, 43 (2019).

\bibitem{Huang2009} Y. Huang, W.N. Polyzou, Phys. Rev. C \textbf{80}, 025503 (2009).

\bibitem{Hummel1994} E. Hummel, J.A. Tjon, Phys. Rev. C \textbf{49}, 21 (1994).

\bibitem{Allen2001} T.W. Allen, W.H. Klink, W.N. Polyzou, Phys. Rev. C \textbf{63},
034002 (2001).

\bibitem{Arenhovel2000} H. Arenhovel, F. Ritz, T. Wilbois, Phys. Rev. C \textbf{61},
034002 (2000).

\bibitem{Phillips20002} D.R. Phillips, Few-Body Syst. Suppl. \textbf{12}, 229
(2000).

\bibitem{Plessas1995} W. Plessas et al., Few-Body Syst. Suppl. \textbf{9}, 429
(1995).

\bibitem{Brodsky1975} S.J. Brodsky, G.R. Farrar, Phys. Rev. D \textbf{11}, 1309
(1975).

\bibitem{Holt2012} R.J. Holt, R. Gilman, Rep. Prog. Phys. \textbf{75}, 086301
(2012).


\end{thebibliography}
\end{document}